# The Application of Virtual Environments and Artificial Intelligence in Higher Education: Experimental Findings in Philosophy Teaching

Adél Vehrer[1]* — Zsolt Pálfalusi[2]

[1] Széchenyi István University, Győr, Hungary, ORCID ID 0000-0001-8884-1020, vehrer.adel@sze.hu
[2] Széchenyi István University, Győr, Hungary, ORCID ID 0000-0002-0144-0038, palfalusi@gmail.com
*Corresponding Author

**Abstract**: *This study explores how virtual environments and artificial intelligence (AI) can enhance university students' learning experiences, with particular attention to the digital preferences of Generation Z. An experiment was conducted at the Apáczai Csere János Faculty of Pedagogy, Humanities, and Social Sciences at Széchenyi István University, where Walter's Cube technology and a trained AI mediator were integrated into the instruction of ten philosophical topics. The curriculum was aligned with the official syllabus and enriched with visual content, quotations, and explanatory texts related to iconic figures in philosophy.*

*A total of 77 first-year undergraduate students from full-time humanities and social sciences programs participated in the study. Following their end-of-semester offline written examination, students voluntarily completed a paper-based, anonymous ten-question test and provided feedback on the method's effectiveness. No sensitive personal data were collected, and the research was conducted with formal approval from the Faculty Dean.*

*Descriptive statistics and inferential tests were applied to evaluate the impact of the virtual environment and AI mediation on learning outcomes. Results indicate that 80% of participants achieved good or excellent final exam grades, and the majority rated the virtual material as highly effective. Qualitative feedback emphasized increased motivation and deeper engagement, attributed to the immersive 3D presentation and interactive AI support.*

*This research contributes to the advancement of digital pedagogy and suggests new directions for applying virtual and AI-based methods in higher education, particularly in disciplines where abstract reasoning and conceptual understanding are central.*

## 1. Introduction

The use of virtual spaces in higher education began in the United States three decades ago when the first VR technologies were experimentally implemented at various universities. The first virtual classrooms appeared around the turn of the millennium. In the mid-2010s, the explosive development of VR technology led to an increasing number of universities adopting VR-based teaching methods, with widespread implementation and true pedagogical integration accelerating over the past 10–15 years. With the advancement of digital technologies, virtual learning environments, particularly 3D virtual reality (VR) applications, have gained significant importance as they enable interactive and experiential learning [1].

The study titled "Discourses of Artificial Intelligence in Higher Education: A Critical Literature Review" comprehensively analyzes how artificial intelligence (AI) is represented in higher education literature. The authors searched for the term "artificial intelligence" in 29 leading higher education journals and examined how articles define and discuss AI.

Two main discourses were identified in the analysis:
1. Discourse of Forced Change: This discourse describes AI as an inevitable change to which higher education institutions must adapt.
2. Discourse of Shifting Authority: In this discourse, AI is presented as a factor that decentralizes the teacher's role, redistributing authority among instructors, machines, companies, and students.

The authors note that the literature contains few clear definitions of AI, and it is rarely treated as an independent research topic. They emphasize the need for new research directions that examine AI's societal impacts, including accountability in AI-mediated practices and how AI influences teaching and learning relationships [2].

It is worth mentioning Zoltán Szűts' study in the Hungarian literature, which examines the role of artificial intelligence in education and teacher training. The author analyzes the opportunities and challenges offered by AI through various scenarios, with particular attention to human-machine collaboration. The study outlines three main scenarios regarding society's relationship with AI:

1. Technology Users: Those who actively and consciously apply AI, enhancing their abilities and efficiency through personalized developments.
2. Drifters: Those who passively and with little awareness use AI, often without fully understanding its operation and effects.
3. Technology Rejecters: Those who reject or lack access to AI, missing out on its benefits and potential growth opportunities.

The author emphasizes that teachers play a key role in teaching the conscious and ethical use of AI. He recommends developing teacher training programs based on human-machine collaboration and fostering digital competencies. Additionally, he highlights the importance of developing critical thinking and problem-solving skills so that students can evaluate and appropriately utilize AI-generated information. The study concludes that integrating AI into education offers numerous opportunities but also raises challenges. Teachers' task is to help students use AI consciously and responsibly, thereby promoting harmonious collaboration between technology and humans [3].

At Széchenyi István University, the methodological integration of the MaxWhere 3D/VR platform into education began in 2017, opening new possibilities for curriculum development and increasing student engagement [4]. The MaxWhere environment was designed to allow users to navigate naturally through three-dimensional spaces while accessing various educational content, including textual, visual, and interactive elements [5]. Additionally, MaxWhere supports collaborative learning by enabling students to work in shared virtual spaces and actively participate in the educational process [6]. One advantage of the platform is that it reduces cognitive load by providing intuitive navigation, contributing to more efficient information processing [7].

Over the past eight years, numerous studies have examined the pedagogical effectiveness of the MaxWhere platform, and the results show that virtual environments improve learning performance and information retention compared to traditional two-dimensional learning environments [8]. In an experimental study, students using MaxWhere achieved significantly better results in recalling visual information than those who studied on traditional e-learning platforms [9]. Since 2018, several studies have confirmed that VR can significantly enhance learning efficiency in higher education, particularly in engineering, healthcare, and natural sciences [10] [8]. Research indicates that this technology not only improves students' learning outcomes but also increases deep learning and motivation compared to traditional online courses [4]. VR technology also aids cognitive development, especially through its positive impact on visual memory and mental rotation abilities. Horváth and Sudár [1] argue that VR-supported education reduces students' cognitive load because three-dimensional environments intuitively facilitate information processing.

VR also plays a significant role in distance education. During the pandemic, many universities were forced to transition to digital education, where VR provided students with the opportunity to learn in a simulated classroom environment, thereby reducing the isolation experienced during online education [5]. VR's application in higher education is particularly important for visualizing complex concepts and theories. In fields such as medicine or engineering, VR enables students to acquire necessary skills and theoretical knowledge in a safe, simulated environment [11]. VR-based learning environments thus open a new dimension in pedagogy, enabling interactive, experiential learning and complementing traditional teaching methods.

In 2024, the development of virtual space-based educational materials expanded with a new element at Széchenyi István University in Győr. The curriculum for the Philosophy course, taught across multiple departments, was developed on the Online Viewing Rooms platform, originally designed for creating online galleries. Walter's Cube, developed by Balázs Faragó and based on Hungarian technology, allows visitors to virtually enter a gallery, walk around, or even purchase artworks as if physically present.

In 2024, it was first proposed that these spaces could also be used for educational content. Compared to 3D/VR educational materials, this interface offers additional functionality, as students can interact with the AI mediator while virtually walking through the space, asking questions in writing or verbally and receiving accurate answers related to the material, assisting them in exam preparation.

## 2. Literature Review

The digital transformation of education has shown significant progress in recent years, particularly with the application of VR technologies. Closely related to this, the rapid development of artificial intelligence (AI) is revolutionizing higher education, providing opportunities to increase the efficiency of educational processes and personalize individual learning experiences.

### 2.1. Virtual-based Learning Environment

Virtual reality has brought about significant transformation in the field of education, especially in higher education and the development of practical skills. VR-based learning facilitates deeper understanding, increases student engagement, and provides opportunities for interactive educational experiences [6].

The application of AI in education is not a new phenomenon; the first intelligent tutoring systems (ITS) appeared as early as the 1960s [12]. By now, AI-based systems are being applied at every stage of the learning process, from adaptive learning to automated evaluation and even predicting student performance [13].

The application of artificial intelligence in higher education can be divided into three main phases:
1. Early Period (1960–1990)
   The first applications of AI appeared in computer-based education. One of the first intelligent tutoring systems, SCHOLAR, was capable of communicating with students in natural language and providing personalized feedback [14].
2. Digital Revolution and E-learning (1990–2010)
   With the rise of the internet, online educational platforms emerged that began applying AI in various areas of education. Learning management systems (LMS), such as Moodle and Blackboard, offered built-in AI functionalities for tracking student performance and supporting adaptive education [15].
3. Generative AI and Intelligent Tutors (2010–present)
   With the development of large language models (LLM), such as GPT-4 and BERT, AI is now capable of complex text processing, answering student questions, and generating personalized educational content [16]. Adaptive learning systems and intelligent tutors are increasingly becoming an integral part of higher education [17].

### 2.2. AI-based Educational Tools and Methods in Higher Education

Artificial intelligence is playing an increasingly important role in higher education, significantly impacting teaching and learning processes. The applications of AI in education include personalized learning systems, automated assessment, the enhancement of teacher-student communication, and the generation and analysis of educational content [18]. Some studies suggest that the use of AI-based tools improves learning outcomes and increases student satisfaction [19]. The technology provides students with the opportunity to experience more interactive and effective learning while offering teachers support in reducing administrative burdens and improving educational methods. In Hungary, a significant proportion of students (87.2%) have already used AI tools in their studies, highlighting the effective use of AI for language learning, data collection, and text generation in education [20]. For example, one study found that the introduction of VR and AI technologies in art education encouraged deep learning, thereby improving students' concentration and creativity [21].

The main areas of artificial intelligence application in higher education are as follows:
1. Adaptive Learning Systems: AI is capable of analyzing student performance in real-time and offering personalized content based on the results. Platforms like Coursera and edX use AI-based analyses to help track student progress and identify knowledge gaps [22].
2. Automated Assessment Systems: AI significantly reduces the workload of instructors by automating the evaluation of essays and tests. For example, the ETS e-rater system uses natural language processing (NLP) to evaluate written assignments [23].
3. Chatbots and Virtual Tutors: AI-based chatbots, such as ChatGPT, are capable of answering student questions, assisting in education, and even providing 24/7 access to information [24].
4. Learning Analytics and Predictive Models: AI-based analytical systems help universities improve the effectiveness of education by predicting student dropout rates and identifying successful learning patterns [25].

2.2.1. *Personalized Learning and Intelligent Tutoring Systems*

AI enables the customization of educational content to meet the needs and abilities of individual learners. AI-based adaptive tutoring systems continuously analyze student performance and adjust the material accordingly [26]. Such systems provide real-time feedback, improving students' motivation and efficiency. For example, Carnegie Learning applied AI in high school mathematics education, tailoring the material to students' individual responses, offering personalized feedback and support, which resulted in better performance among students [27].

According to recent studies, generative AI, such as ChatGPT, can also be an effective tool for personalized learning, particularly in essay writing and creative problem-solving [28]. These tools provide an interactive learning experience, offering real-time feedback to students so they can receive timely input and develop a deeper understanding of the material.

One of the most important applications of AI is the development of personalized learning systems. Intelligent learning systems are capable of adapting to students' individual abilities and learning pace, thereby optimizing learning outcomes [26]. Studies show that the use of such systems improves student performance, especially for disadvantaged learners, for whom AI can provide equal access to quality education [29].

## 3. Virtual Teaching Assistants and Chatbots

AI-based virtual assistants and chatbots provide real-time support to students, answering their questions and guiding them through the learning process. These virtual assistants not only respond to students' queries but also offer relevant explanations, suggestions, or resources. This is particularly useful in e-learning systems where students progress independently through the material and require immediate support. The use of such chatbots and virtual assistants contributes to creating a personalized learning experience, as they can take into account students' individual needs and preferences. These tools can also reduce teachers' workload by handling repetitive questions and administrative tasks [30].

## 4. Teacher-Student Collaboration and Communication

AI-based chatbots and intelligent tutors are increasingly being integrated into the operations of educational institutions. According to DiCerbo [31], the Khan Academy's AI-based tutor, Khanmigo, opens up new possibilities for customizing education across various subjects while also assisting teachers with lesson planning and data analysis. These tools enable students to receive direct feedback during their learning process, significantly improving educational efficiency [32]. These tools also facilitate communication and collaboration between teachers and students.

AI-based chatbots and virtual teaching assistants are capable of answering simpler questions, recommending learning materials, and tracking students' progress. Some studies suggest that AI-based education can increase learning efficiency and is useful for the early identification of students' academic difficulties [26].

## 5. Use of AI in Domestic Higher Education

Bokor [33] examined the opportunities and limitations of applying artificial intelligence in education. The study highlights that integrating AI into education offers numerous possibilities for enriching the learning experience and enhancing pedagogical efficiency. However, it is essential to thoroughly examine ethical and societal issues and adopt a technology-invariant approach to ensure that AI truly creates value in education. The study identifies four main areas where AI can be effectively applied:
1. Intelligent Tutoring Systems (ITS): These systems can provide personalized learning experiences, adapting to the individual needs and abilities of learners.
2. Virtual Learning Environments: AI supports virtual and augmented reality-based educational platforms, offering interactive and immersive learning experiences.
3. Automated Assessment Systems: The use of AI allows for the rapid and objective evaluation of student performance, such as the automatic grading of essays or short-answer questions.
4. Student Support and Counseling: AI-based chatbots and virtual assistants can provide 24/7 support to students, answering their questions and assisting them throughout the learning process.

According to Jäckel and Garai-Fodor [34], the majority of students at Óbuda University are open to using AI during their studies, particularly for information retrieval and utilizing learning aids. Students primarily use AI for data collection, information search, and learning aids. The study points out that while Generation Z is open to using AI in higher education, there is a need for appropriate guidance and training to ensure they use these technologies consciously and responsibly during their studies.

Rajki et al. [20] also note that artificial intelligence is increasingly being integrated into higher education, particularly in supporting student learning processes. The research focused on how extensively students use ChatGPT and similar generative AI tools. The study involved 1,027 university students, mainly from humanities, social sciences, and education programs. It found that 87.2% of students had already used some form of AI tool during their studies. The most popular applications were Duolingo (61.3%) and ChatGPT (40.9%), indicating that AI is predominantly used for language learning and text generation. Additionally, 39.4% of participants used AI to create essays or theses, raising concerns about plagiarism and academic integrity. The main barriers included lack of knowledge (48.0%), ethical issues (41.6%), and data privacy concerns (30.2%).

Marciniak and Baksa [35] examine the impact of generative artificial intelligence tools (GMI tools), such as ChatGPT or Google Bard, on higher education, with particular attention to the humanities and social sciences. The authors emphasize that key stakeholders in higher education must engage in transparent communication about the expected impacts of GMI tools to quickly establish a culture of proper usage.

## 6. The Use of Virtual Spaces in Arts and Humanities Education

The growing application of VR and AI is also evident in the fields of arts and humanities, as well as social sciences. For example, historical and archaeological reconstructions enable students to interactively study the past, while in psychology and sociology, VR can be used to examine simulated social interactions and behavioral patterns [36].

Virtual spaces, such as the metaverse (augmented reality (AR), virtual reality (VR), and traditional internet platforms), offer new opportunities in education, particularly in facilitating deeper learning. The use of VR in learning environments allows students to interact with course materials in an engaging way, increasing motivation and deepening the learning experience [21]. VR options provide students with the ability to practice philosophical argumentation and debate in a virtual environment, which promotes the development of critical thinking skills.

According to [32], the use of avatars and interaction within virtual reality offers students the opportunity to actively participate in the learning process, which is particularly beneficial for understanding complex philosophical topics. These tools provide opportunities for innovation in educational methods, especially in philosophy education, where understanding abstract concepts can pose challenges for students.

The application of AI tools enables universities to create more supportive and inclusive learning environments for all students. As a result, higher education can become more personalized, efficient, and accessible [18]. In the context of philosophy education, studies have already been conducted on how artificial intelligence can be integrated into teaching philosophy and developing critical thinking. AI tools allow students to gain a deeper understanding of complex philosophical concepts and encourage them to engage in independent, critical thinking. Additionally, they enable students to explore different perspectives on complex issues. Furthermore, AI algorithms are capable of analyzing patterns of thought, providing feedback on logical errors or cognitive biases, and assisting students in improving their reasoning skills [37].

The integration of AI and virtual spaces in education not only enriches the learning experience but also contributes to the development of students' digital competencies. The use of digital tools and AI in education fosters students' critical thinking, creativity, and problem-solving abilities, which are essential in the 21st-century job market. Similarly, the combination of VR and artificial intelligence transforms traditional teaching methods by offering personalized and captivating educational experiences that can reshape how students interact with course materials [38].

## 7. Future Perspectives on AI Applications

The role of artificial intelligence in higher education is expected to grow further in the following areas:
1. AI Combined with Virtual Reality: The joint application of AI and VR/AR technologies enables experiential learning [39].

2. Development of Generative AI: AI may expand its capabilities in generating educational content and tutoring systems [16].
3. Ethical AI Development: Ensuring transparent and responsible AI systems will be crucial in higher education [40].

The application of AI in education has the potential to revolutionize teaching and learning processes, particularly in personalized learning, automated assessment, and teacher-student collaboration. New technologies, such as VR and AI-based tutors, can further enhance the quality and accessibility of education. However, due to the ethical and regulatory challenges posed by AI, it is important to ensure responsible and transparent usage that supports equitable and effective educational environments.

In philosophy education, the application of VR and AI is particularly useful for understanding abstract concepts and theories. For example, students can use virtual experiences to explore various philosophical movements and thinkers, leading to deeper comprehension and engagement. AI-based tutors can provide personalized support to students, helping them understand and analyze complex philosophical questions.

## 8. Methodology and results

Within the framework of the research, 77 students explored the course material in ten 3D "rooms" while interacting with an AI mediator. At the end of the experiment, participants completed a ten-question anonymous test and provided feedback on the effectiveness of the method. Descriptive statistics were applied to evaluate the data and examine how the combination of virtual technology and the AI mediator influenced learning outcomes.

The sample consisted of first-year undergraduate students enrolled in full-time programs in the humanities and social sciences. Data collection was conducted immediately following the end-of-semester offline written examination. Participation was entirely voluntary, and students completed the paper-based questionnaire on-site without incentives. No sensitive personal data were collected, and informed consent was obtained prior to participation. The study was carried out with formal approval from the Faculty Dean, in accordance with institutional guidelines.

**Figure 1.** Virtual gallery environment used in the experiment, featuring Immanuel Kant and AI-supported interaction. Source: Screenshot by the author.

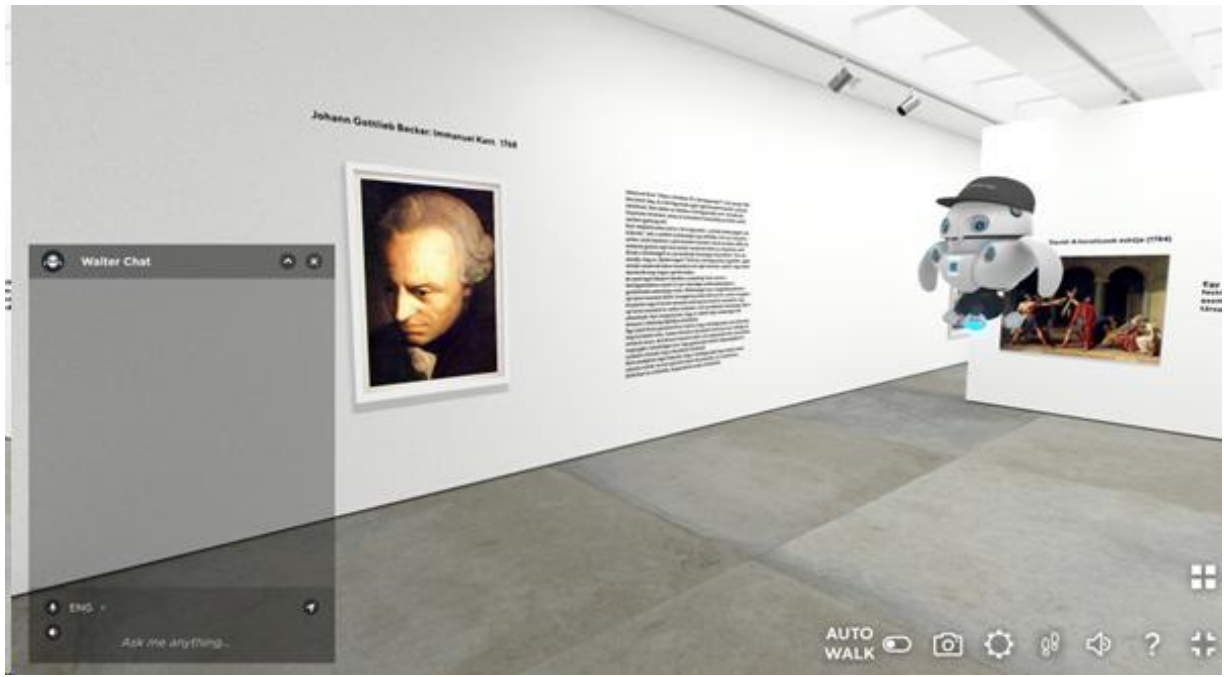

Number of participants: 77 students
Learning environment: 10 different interactive 3D "rooms"
Learning support: Interaction with an AI mediator
Testing: 10-question anonymous final test

**Table 1** Descriptive Statistics

| Variable | Value |
|---|---|
| Total number of participants | 77 |
| Students who rated the material as excellent | 54 (70.1%) |
| Students who rated the material as good | 18 (23.4%) |
| Students willing to apply the technology in other subjects | 76 (98.7%) |
| Proportion of students achieving a good or excellent grade on the final test | 80% |

Source: Author's own data.

In the analysis conducted using SPSS 25.0, the following descriptive statistics were applied:
− Frequency analysis: To examine the distribution of participants' evaluations, intentions, and test results.
− Relative frequency (%): To present the percentage distribution of responses.

**Figure 2.** The Nietzsche section of the virtual exhibition with Hungarian explanatory text and AI interface.
Source: Screenshot by the authors.

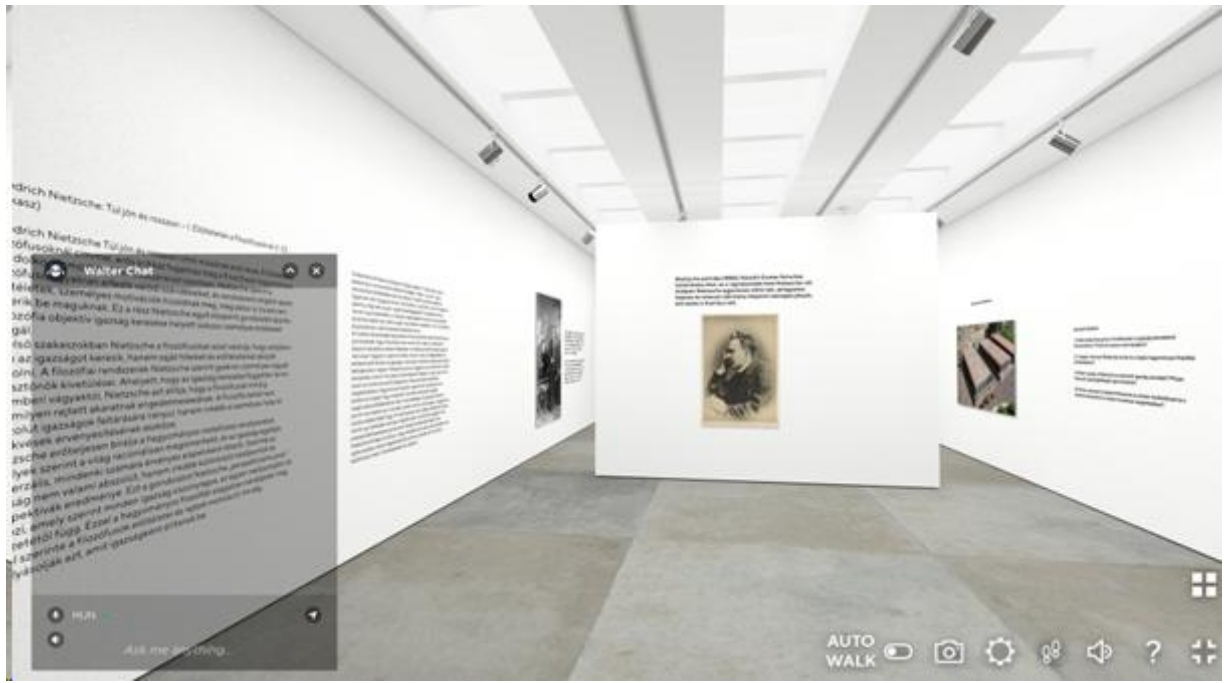

*8.1. Supplement: Inferential Statistical Analysis*

Beyond descriptive statistics, inferential statistical tests were also applied to determine whether significant relationships existed between student feedback and test results.

*8.1.1. Chi-square test (χ²-test) – Comparison of categories*

Research question: Is there a significant relationship between whether someone rated the material as excellent or good and whether they would be willing to use it in other subjects?

**Table 2** Contingency table based on data

|  | **Would Use in Other Subjects** | **Would Not Use** | **Total** |
|---|---|---|---|
| Rated as Excellent | e.g., 53 | 1 | 54 |
| Rated as Good | e.g., 17 | 1 | 18 |
| Total | 70 | 2 | 72 |

**Source:** Author's own data.

Assumption: 76 participants indicated willingness to use the technology, so most positive evaluators were included.

*8.1.2. Binomial test or Z-test for proportions*

Research question: Did a significantly higher proportion of students achieve good or excellent grades compared to what would be expected by chance?
Test results:
Achieved good or excellent grade: 80%
Expected proportion (e.g., assumed 50% without a control group): $p_0 = 0.5$
Result:

p < 0.001 → Significantly more students achieved good or excellent results than would be expected by chance.

*8.2. Summary*

The applied inferential statistical tests confirmed that:
Positive evaluations were significantly associated with the intention to apply the material in other subjects.
Participants' test results exceeded the expected level of random performance, reinforcing the effectiveness of the 3D presentation and AI mediator.

**9. Conclusion**

Based on the feedback, the overwhelming majority of students evaluated the learning experience positively. The combined effect of the 3D presentation and the AI mediator increased learning motivation and deepened understanding. Qualitative feedback frequently highlighted sustained attention, easier comprehension, and support for independent learning.

The results of the experiment highlight that the combination of virtual space and AI can significantly enhance students' preparation and lead to better exam results. This approach is particularly relevant for Generation Z, who prefer digital, interactive learning environments. Our research contributes to the development of digital teaching methods and points to new directions for further experiments in higher education, including the applicability of the method to other subjects and the potential for future technological advancements.

**Transparency**

The authors confirm that the manuscript is an honest, accurate, and transparent account of the study; that no vital features of the study have been omitted; and that any discrepancies from the study as planned have been explained. This study followed all ethical practices during writing.